%% file: main.tex
\documentclass[%reprint,
superscriptaddress,
 amsmath,amssymb,
 aps,
prl,
twocolumn,
]{revtex4-2}
\newcommand{\thor}{$^{229}$Th}
\newcommand{\regthor}{$^{232}$Th}
\newcommand{\lisaf}{LiSrAlF$_6$}

\usepackage[dvipsnames]{xcolor}
\definecolor{ricky}{cmyk}{0, 0.7808, 0.4429, 0.1412}

\usepackage{braket}
\usepackage{graphicx}
\usepackage[colorlinks=true, allcolors=blue]{hyperref}
\usepackage{gensymb}

\usepackage{upgreek}

\usepackage{float}  
\usepackage{xcolor}

\usepackage{braket}
\usepackage{graphicx}

\usepackage[colorlinks=true, allcolors=blue]{hyperref}
\usepackage[version=3]{mhchem}

\usepackage{upgreek}

\begin{document}

\title{Laser excitation of the \thor\ nuclear isomeric transition in a solid-state host}

\author{R. Elwell}
\affiliation{Department of Physics and Astronomy, University of California, Los Angeles, CA 90095, USA}
\author{Christian Schneider}
\affiliation{Department of Physics and Astronomy, University of California, Los Angeles, CA 90095, USA}
\author{Justin Jeet}
\affiliation{Department of Physics and Astronomy, University of California, Los Angeles, CA 90095, USA}
\author{J. E. S. Terhune}
\affiliation{Department of Physics and Astronomy, University of California, Los Angeles, CA 90095, USA}
\author{H. W. T. Morgan}
\affiliation{Department of Chemistry and Biochemistry, University of California, Los Angeles, Los Angeles, CA 90095, USA}
\author{A. N. Alexandrova}
\affiliation{Department of Chemistry and Biochemistry, University of California, Los Angeles, Los Angeles, CA 90095, USA}
\author{H. B. Tran Tan}
\affiliation{Department of Physics, University of Nevada, Reno, Nevada 89557, USA}
\affiliation{Los Alamos National Laboratory, P.O. Box 1663, Los Alamos, New Mexico 87545, USA} 
\author{Andrei Derevianko}
\affiliation{Department of Physics, University of Nevada, Reno, Nevada 89557, USA}
\author{Eric R. Hudson}
\affiliation{Department of Physics and Astronomy, University of California, Los Angeles, CA 90095, USA}
\affiliation{Challenge Institute for Quantum Computation, University of California Los Angeles, Los Angeles, CA, USA}
\affiliation{Center for Quantum Science and Engineering, University of California Los Angeles, Los Angeles, CA, USA}
\date{\today} % Leave empty to omit a date

\begin{abstract}
LiSrAlF$_6$ crystals doped with \thor\ are used in a laser-based search for the nuclear isomeric transition. 
Two spectroscopic features near the nuclear transition energy are observed.
The first is a broad excitation feature that produces red-shifted fluorescence that decays with a timescale of a few seconds.
The second is a narrow, laser-linewidth-limited spectral feature at $148.38219(4)_{\textrm{stat}}(20)_{\textrm{sys}}$~nm ($2020407.3(5)_{\textrm{stat}}(30)_{\textrm{sys}}$~GHz)
that decays with a lifetime of $568(13)_{\textrm{stat}}(20)_{\textrm{sys}}$~s. 
This feature is assigned to the excitation of the \thor\ nuclear isomeric state, whose energy is found to be $8.355733(2)_{\textrm{stat}}(10)_{\textrm{sys}}$~eV in \thor:\lisaf.
\end{abstract}

\maketitle

The nuclear isomeric state in the \thor\ nucleus, described by the Nilsson quantum numbers $(3/2)^+\left[631\right]$ has the lowest energy of all known nuclear excited states.
This extraordinary property, coupled with its expected long lifetime, should allow access to a number of
novel applications, including construction of an optical nuclear clock that may be more robust~\cite{Rellergert2010a} than and/or  outperform~\cite{PeikTamm2003,Campbell2012} current technology.
It is also expected to allow the most sensitive test to date of the  variation of the fundamental constants~\cite{Rellergert2010a,Flambaum2006,Litvinova2009}.

While recent work has greatly improved the knowledge of the isomeric state energy~\cite{vonderWense2016,Seiferle2019,Sikorsky2020}, to realize the aforementioned goals requires an energy measurement with laser spectroscopic precision of the isomeric transition, $(3/2)^+\left[631\right] \leftarrow (5/2)^+\left[633\right]$.
It was proposed~\cite{Rellergert2010a} that due to the optical M\"{o}{\ss}bauer effect, a high-bandgap crystal doped with \thor\ might provide an attractive platform for performing this spectroscopy. 
As a result, considerable effort has been put towards developing high-quality \thor-doped crystals, both by traditional methods~\cite{Jeet2015, Beeks2023} and using implantation at a radioactive beamline~\cite{Kraemer2022a}. 
This led to an important experiment at ISOLDE~\cite{Kraemer2022a}, which established that detection of a long-lived nuclear decay ($\tau \sim 1000$~s) is possible in a crystalline host and that the wavelength of the isomeric transition was 148.7(4)~nm -- parentheses denotes the 68\% confidence interval.

Here, we report the results of a laser-based search for the \thor\ isomeric nuclear transition in \thor-doped LiSrAlF$_6$ crystals. 
Using a vacuum-ultraviolet (VUV) laser system~\cite{JeetThesis2018}, we have searched for the transition between 147.43~nm-182.52~nm (6.793~eV-8.410~eV).
We observe wideband fluorescence from \thor-doped crystals in the spectral region identified by the ISOLDE experiment~\cite{Kraemer2022a}, while undoped and \regthor-doped crystals do not exhibit similar fluorescence. 
This fluorescence appears to possess multiple time scales that are of order a few seconds.
We also observe a narrow, laser-linewidth-limited spectral feature at $148.38219(4)_{\textrm{stat}}(20)_{\textrm{sys}}$~nm, which decays with a lifetime of $568(13)_{\textrm{stat}}(20)_{\textrm{sys}}$~s.
This feature does not appear in a \regthor{}-doped LiSrAlF$_6$ crystal. 

Very recently, using the approach of Ref.~\cite{Rellergert2010a}, a narrow spectroscopic feature was observed in the laser spectroscopy of \thor{}-doped CaF$_2$ crystals at 148.3821(5)~nm~\cite{Tiedau2024}.
Since this feature did not appear in a \regthor{}-doped CaF$_2$ crystal it was assigned to the \thor{} isomeric transition.
Such assignment relies on the chemical identity of \regthor{}- and \thor{}-doped crystals, which, given that the radioactivity of \thor{} may lead to formation of radiation-induced color centers in \thor{}-doped crystals, is not guaranteed -- especially given the history of the search for the isomeric transition~\cite{Irwin1997,Shaw1999}. 
However, given that the feature is observed in both \thor{}-doped CaF$_2$ and LiSrAlF$_6$ crystals under different experimental conditions, it appears safe to assign it to the \thor{} nuclear isomeric transition. 

To interpret these results, we present \emph{ab initio} crystal structure calculations that suggest the observed short-time fluorescence could be due to
coupling of the \thor{} nucleus to the electronic and phononic degrees of freedom of the crystal.
This coupling may also explain why only a fraction of the \thor{} nuclei doped into the crystal appear to contribute to the observed narrow, isomeric transition.

The experiments reported here were conducted via monitoring the resulting fluorescence from crystals following illumination with a VUV laser.
Four different VUV transparent LiSrAlF$_6$ crystals were used in these experiments, all with dimensions $\approx$~3~mm~$\times$~3~mm~$\times$~10~mm. 
Two were doped with \thor\ (crystal 2.2 with \thor\ density $\rho_{229} \approx 5(10^{16})$~cm$^{-3}$ and crystal 3.1 with $\rho_{229} \approx 5(10^{15})$~cm$^{-3}$), 
one was doped with \regthor\ ($\rho_{232} \approx 1(10^{16})$~cm$^{-3}$) to understand any fluorescence resulting from the chemistry of Th in LiSrAlF$_6$, and one was an undoped LiSrAlF$_6$ crystal to understand any fluorescence due to intrinsic properties of the host material. 
\begin{figure*}[t!]
    \centering
    \includegraphics[width=1\textwidth]{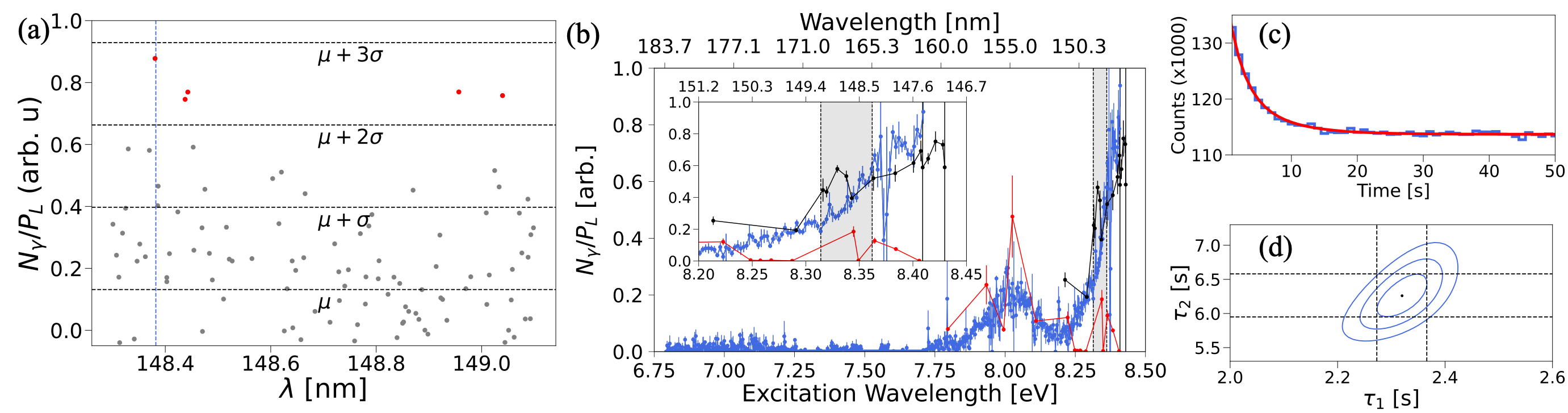}
    \caption{(a)~Long timescale fluorescence from \thor:\lisaf\ crystal 2.2 normalized by laser power after $\approx~ 2000$~s illumination shown over the 68\% confidence interval of the ISOLDE experiment~\cite{Kraemer2022a}.
    Dotted black lines denote the mean ($\mu$) and mean plus multiples of the standard deviation ($\sigma$).
    Though the expected signal is roughly two-orders of magnitude larger than what is observed, points above $\mu + 2\sigma$ were denoted as potential detection events and slated for further study.
    The vertical blue dashed line denotes the location of the narrow feature observed in Fig.~\ref{fig:NarrowLine}(a).
    (b)~Short timescale fluorescence normalized by laser power vs. excitation energy in \thor\ (2.2 in blue, 3.1 in black) and \regthor\ (red) \lisaf~crystals. The inset shows the region near the ISOLDE 68\% confidence region (shaded). 
    Each of the blue data points required roughly 1~hr to collect.
    The drop in signal near 8.37~eV is due to a resonance in the Xe gas used for generating the VUV.
    (c) Time binned histogram of the short time fluorescence from \thor\ crystal 2.2 fitted with a two timescale exponential decay model.
    (d) The likelihood contours of the two timescales found from the data in (c). Each contour represents the [68\%, 95\%, 99.7\%] confidence regions around $\chi^2_{min}$.}
    \label{fig:panel_plot}
\end{figure*}

VUV radiation was produced via resonance-enhanced four-wave mixing of two pulsed dye lasers in Xe gas~\cite{JeetThesis2018, Berdah1996}. 
The frequency of the first pulsed dye laser, $\omega_u$,
was locked to the $5p^{6 ~1}S_0~\rightarrow~5p^5\left(^2P^{\circ}_{3/2} \right) 6p~^2\left[1/2\right]_0$ two-photon transition of Xe. 
The frequency of the second pulsed dye laser, $\omega_v$, was scanned 
to produce VUV radiation in the Xe cell given by the difference mixing relation $\omega = 2\omega_{u} - \omega_v$. 
All three laser beams then impinge on an off-axis MgF$_2$ lens, whose chromatic dispersion is used with downstream pinholes to spatially filter the VUV beam and pass it towards the crystal chamber. 
The laser system deliveries 30 pulses per second to the crystal with a typical VUV pulse energy of $\sim1$ $\mu$J/pulse (see Ref.~\cite{Jeet2015,JeetThesis2018} for details).

The crystal under study is held in a vacuum chamber consisting of a crystal mount, two VUV-sensitive photo-multiplier tubes (Hamamatsu R6835/R6836), and a pneumatic shutter system to protect the PMTs from direct exposure to the VUV laser.
The PMTs are operated in a cathode-grounded configuration, and their output waveforms recorded by a 1~Gs/s waveform digitizer (CAEN DT5751) for subsequent post-processing. 
The VUV laser beam terminates on a custom VUV energy detector.
The crystal chamber is maintained with an Ar atmosphere at a pressure of $\sim$10$^{-2}$~mbar to provide high VUV transmission while minimizing browning of optics due to hydrocarbon deposition~\cite{JeetThesis2018}.

Using this system, an experiment was performed from May 2016 to July 2019 that was designed to search 
for crystal fluorescence with a lifetime of $\approx 1000$~s, as would be expected for radiative relaxation of the isomeric level in a \thor:\lisaf\ crystals~\cite{Tkalya2015}.
For this experiment, crystal 2.2 was illuminated by VUV radiation for roughly 2000~s, the shutters were opened within 0.5~s, and any resulting fluorescence collected by the PMTs for roughly 1000~s. 

For each data point constituting 2000~s of illumination, the VUV laser frequency was swept 8 times over $\sim$160 GHz. 
Each subsequent data point was then overlapped in frequency by $\sim$40 GHz with the previous one.
This procedure was employed to minimize the probability of failing to illuminate the isomeric transition due to, e.g., multi-mode behavior of the pulsed dye lasers without requiring impractical data collection times.

Figure~\ref{fig:panel_plot}(a) shows the number of detected photons after background correction ($N_\gamma$) normalized by the VUV laser power ($P_L$) over the ISOLDE-identified 68\% confidence interval~\cite{Kraemer2022a}.
The expected signal, assuming that all doped \thor\ nuclei are optically addressable, is roughly two orders of magnitude higher than the observed level.
However, by analyzing the distribution of the detected photons, points deviating from the mean by more than two standard deviations were flagged for further study, as described later. 

In contrast, a clear excess of detected photons can be seen in Fig.~\ref{fig:panel_plot}(b), which shows $N_\gamma/P_L$ in the first 50~s after the laser is extinguished as a function of excitation wavelength.
This fluorescence appears largest near the spectral region identified by the ISOLDE experiment~\cite{Kraemer2022a}.
Analysis suggests that this fluorescence possesses multiple timescales and is reasonably fit by a two-lifetime model as shown in Fig.~\ref{fig:panel_plot}(c); the likelihood contours for these two lifetimes are shown in Fig.~\ref{fig:panel_plot}(d).

Following this observation, several additional experiments were performed to ascertain the origin of the short-timescale fluorescence.
For these experiments, the crystals were illuminated by the VUV laser for 60~s, the shutters opened in 100~ms, and the resulting fluorescence observed for roughly 60~s. 
First, a second \thor:\lisaf\ crystal (batch 3.1) was investigated to confirm the existence of the spectral feature. 
The resulting fluorescence is shown as the black line in Fig.~\ref{fig:panel_plot}(b) and follows closely the observed fluorescence in crystal 2.2; the fluorescence exhibits a two-timescale behavior that is indistinguishable from that of crystal 2.2.
Following these measurements, the experiment was repeated with a \regthor:\lisaf\ crystal (red line).
As can be seen in Fig.~\ref{fig:panel_plot}(b), the \regthor:\lisaf\ and \thor:\lisaf\ crystals yield a similar spectroscopic feature around 8~eV, but only the \thor:\lisaf\ crystals yield a significant signal near the nuclear transition.
Assuming this feature arises from nuclear excitation, we estimate from the known \thor\ density in these crystals~\cite{Jeet2015} and the detection efficiency of the system that only a portion of order 1~ppm of the doped \thor\ atoms participate in this short-timescale process. 
Other explanations, such a radiation-induced color centers cannot be ruled out. 

\begin{figure}
    \centering
    \includegraphics[width=0.48\textwidth]{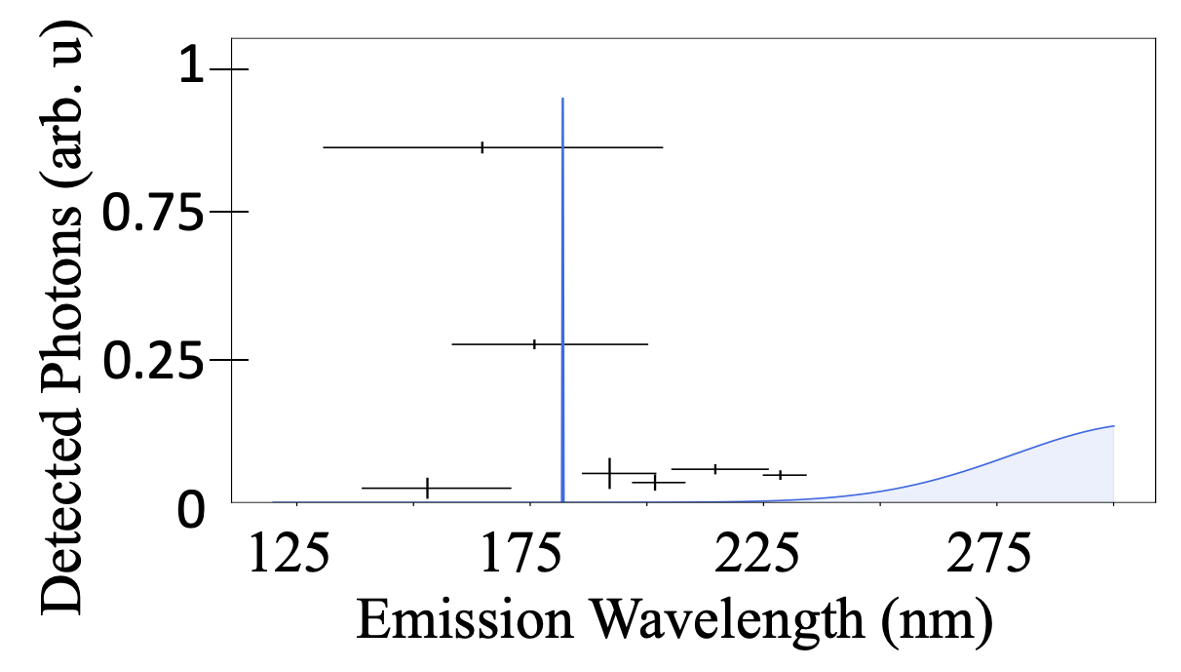}
    \caption{Fitted spectrum for the fluorescence from crystal 2.2 under excitation at 147.4 nm using a Gaussian deconvultion (see SI). 
    The black points represent the recorded photon numbers; the vertical line represents the statistical uncertainty while the horizontal line represents the full width at half maximum (FWHM) of the transfer function of that particular filter configuration. 
    Each point is centered at the peak of the transfer function.}
    \label{fig:filter_trans}
\end{figure}

This data suggests the observed short-timescale fluorescence could have a nuclear origin and it is therefore interesting to determine the wavelength of emission. 
However, the number of emitted photons was sufficiently small ($\sim10^4$ photons into $4\pi$) that it could not be analyzed with available monochromators. 
Therefore, in order to determine the spectrum of the fluorescence emitted from the highest activity \thor:\lisaf\ crystal (crystal 2.2), the transmission of the fluorescence was measured through a set of band-pass filters. 
An imaging system was used to collimate the collected light before passing through the filters in order to eliminate angle effects~\cite{Lofdahl2011}.
The filtered light was then focused onto a photomultiplier tube for detection. 
The VUV laser wavelength was chosen to be $\approx 147.4$~nm to maximize the amount of fluorescence.
The resulting fluorescence through various filter configurations is plotted in Fig.~\ref{fig:filter_trans}, where the horizontal axis is the central transmission wavelength of each configuration and the horizontal `error bar' denotes the FWHM of the filter transmission (see SI).
As can be seen, the fluorescence is clearly in the VUV region as it primarily transmits through only two filters nominally centered at 154~nm and 180~nm. 
To better determine the wavelength of the emitted light, a Gaussian deconvolution procedure~\cite{Major2020} was used to fit a model spectrum to the data (see SI).
This analysis is most consistent with a narrow emission at $\approx 182$~nm with a long-tailed background centered red of $280$~nm.
The long-tail in the UV is expected and has previously been understood as due to fluorescence induced by radioactive decay of the \thor\ in the crystal (see Fig. 2(b) of Ref.~\cite{Jeet2015}).
As will be discussed later, the narrow VUV emission could be attributed to the isomeric state quenched by excitation of a defect electron-hole pair in the crystal. 

\begin{figure}[t!]
    \centering
    \includegraphics[width = 0.47\textwidth]{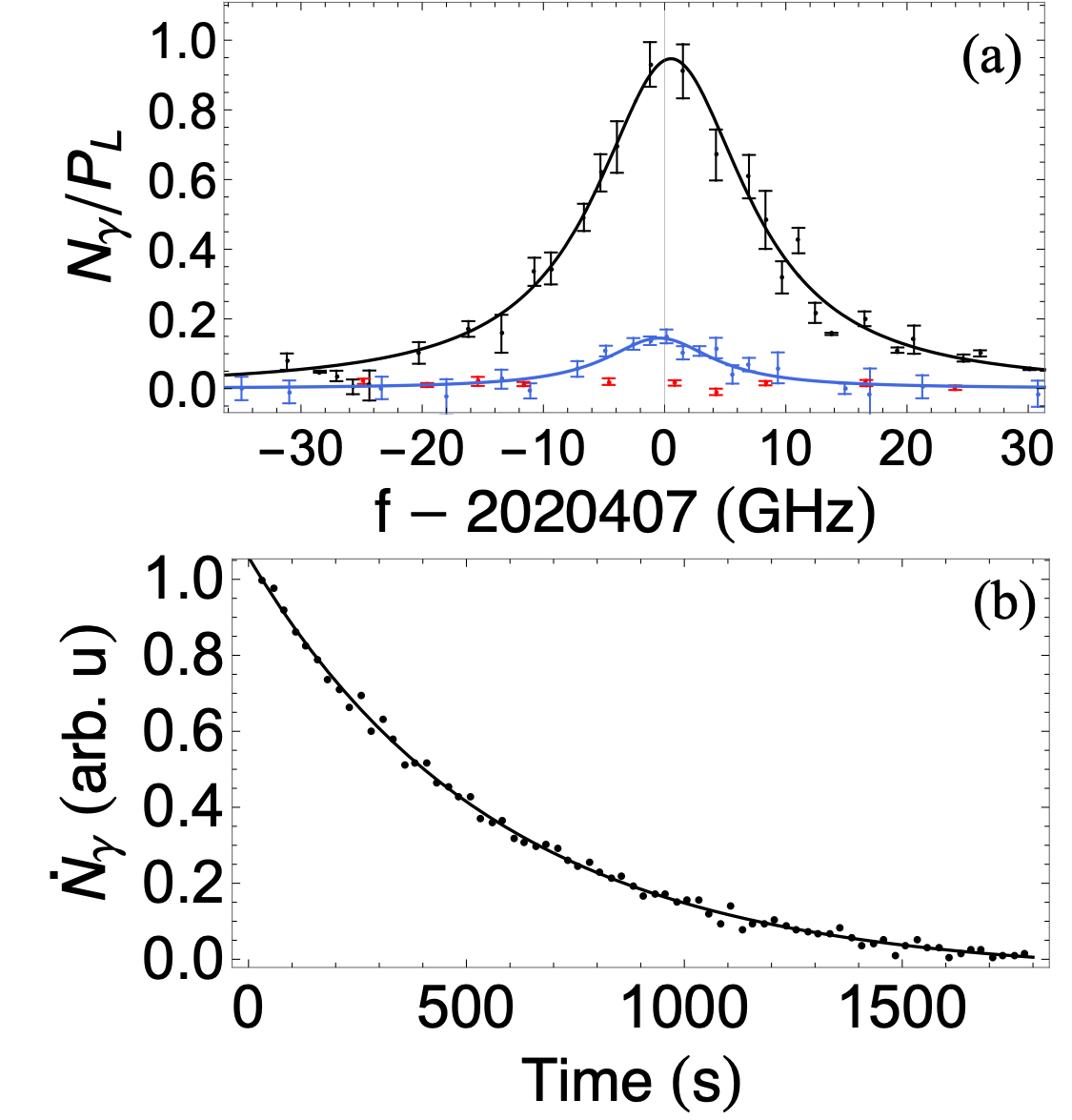}
    \caption{(a) Spectrum of the \thor{} isomeric transition. The total power-normalized photon count in 1800~s following illumination is plotted versus excitation wavelength for crystal 2.2 (blue), 3.1 (black) and a \regthor:\lisaf{} crystal (red).
    The solid lines represent fits of a Lorentzian lineshape, which result in statistically identical center frequencies. 
    The linewidth of the fitted Lorentzians are also statistically identical and consistent with the linewidth of the VUV laser system, indicating a narrow line. 
    (b) Observed photon count rate versus time from crystal 3.1 alongside an decaying exponential fit which determines the isomeric state lifetime within the crystal to be $\tau = 568(13)_{\textrm{stat}}(20)_{\textrm{sys}}$~s. A statistically identical lifetime is observed in crystal 2.2.}
    \label{fig:NarrowLine}
\end{figure}

Following this study of the short-timescale fluorescence, we returned to the study of the long-timescale fluorescence.
The ISOLDE experiment had found that only a few percent of the \thor\ atoms implanted into the crystal contributed to the radiative signal~\cite{Kraemer2022a}.
This suggests that some lattice sites are quenched by coupling to the electronic structure of the crystal. 
Therefore, we began a new search, optimized for detection of the isomeric transition, assuming only a few percent of the doped \thor\ atoms contributed.
This search was performed over the spectral regions within the ISOLDE region that showed an excess of photon counts relative to their neighbors (see Fig.~\ref{fig:panel_plot}(a)).
The results of that optimized search near a region of excess photons counts in our 2019 data (see dashed blue line 
in Fig.~\ref{fig:panel_plot}(a)), are shown in Fig~\ref{fig:NarrowLine}(a).
The total power-normalized, long-timescale fluorescence observed in this region reveals a narrow spectral feature at $148.38219(4)_{\textrm{stat}}(20)_{\textrm{sys}}$~nm in both crystals 2.2 and 3.1; the primary source of systematic uncertainty is wavemeter calibration (see SI).
The full width at half maximum (FWHM) of this feature is roughly 15~GHz and in agreement with a measurement of the laser linewidth.
The central wavelength of this feature is identical, within error, in both crystals 2.2 and 3.1 and to that observed in Ref.~\cite{Tiedau2024}.
It is, thus, attributed to the excitation of the \thor\ isomeric state.
From this data, we estimate that approximately 1\% (50\%) of the doped \thor{} nuclei in crystal 2.2 (3.1) are able to be excited on the nuclear isomeric transition.

The measured isomeric decay lifetime is $\tau = 568(13)_{\textrm{stat}}(20)_{\textrm{sys}}$~s as shown in Fig.\ref{fig:NarrowLine}(b), where the systematic error is estimated based on long-term drifts in the PMTs background count rates. 
The LiSrAlF$_6$ index of refraction is not well-characterized in the VUV but can be estimated as 1.485~\cite{JeetThesis2018}, leading to an estimate of the isomeric lifetime in vacuum of $\tau_{is} \approx 1860(43)_{\textrm{stat}}(66)_{\textrm{sys}}$~s. 
This inferred lifetime is similar to that observed in \thor:CaF$_2$~\cite{Tiedau2024} (1740(50)~s), especially given uncertainty in the indices of refraction of these crystals in the VUV, and in agreement with that predicted in Ref.~\cite{Tkalya2015}.

\begin{figure}[t!]
    \centering
    \includegraphics[width = 0.48\textwidth]{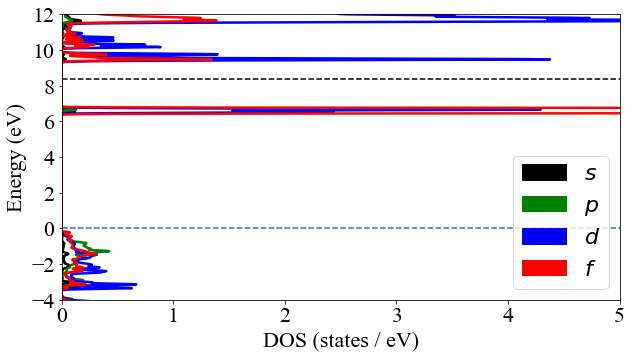}
    \caption{Thorium PDOS for \thor:LiSrAlF$_6$. The dashed black line represents the energy of the \thor\ nuclear excited state relative to the top of the valence band.}
    \label{fig:DOS}
\end{figure}

To better understand these results, we performed density functional theory (DFT) calculations using the modified Becke-Johnson~\cite{RN489,RN490} functional to estimate the electronic properties of the \thor-doped crystals (see SI).
Fig.~\ref{fig:DOS} shows the projected density of states (PDOS) for Th in Th:LiSrAlF$_6$, in a $3 \times 3 \times 2$ supercell containing one Th atom and 2 interstitial F atoms for charge balancing; this is the lowest-energy configuration in our calculations (see SI). 
We find a manifold of the Th $5f$ states at 6.8 eV (182~nm) inside the insulator bandgap ($d_i$ states). 
This is consistent with the observed fluorescence wavelength in Fig.~\ref{fig:filter_trans}, indicating the possibility of nuclear decay via excitation of a valence-band electron to a $d_i$ state.
Subsequent photon emission via the recombination of the $d_i$ state with the hole, which has moved to the top of the valence band on ps timescales typical for non-radiative relaxation of hot holes~\cite{Woerner1993}, would produce the observed, red-shifted 182~nm photons.

Such coupling of the electron and nuclear sub-systems is mediated by the hyperfine interaction (HFI). 
In the absence of the HFI, the eigenstates of the system are the product states $|\mathrm{el} \rangle |\mathrm{nuc} \rangle$; the HFI leads to admixtures of the excited nuclear state $|e\rangle$ with the product states involving the ground $|g\rangle$ nuclear state. 
In the first part of the process, the laser excites the occupied valence band state $\approx |\Omega\rangle |g \rangle$ to a particle-hole eigenstate $| \Psi_L \rangle =a_{d_i}^\dagger a_h |\Omega\rangle |g \rangle + |\delta \Phi_\mathrm{el}\rangle |e \rangle$ resonant with the laser frequency (here we use the creation and annihilation operators $a^\dagger$ and $a$; $|\Omega\rangle$ is the valence band quasi-vacuum state). The HFI-mediated admixture $|\delta \Phi_\mathrm{el}\rangle |e \rangle$ can be substantial due to a contribution of the  $a_{d_i}^\dagger a_{h_r} |\Omega\rangle$ states resonant with the nuclear transition. It also contains the $|\Omega\rangle |e \rangle$ state.  Such laser excitation exhibits a step-like character of Fig.~\ref{fig:panel_plot}(b).
The components of this coherently excited state decay due to a combination of vibronic couplings to the crystal on a ps timescale and radiative particle-hole recombination on a ns timescale. While the majority of the decays result in a photon emission on the ns timescale, some population may evolve into the state $|\Omega\rangle|e\rangle$.
The resulting $|\Omega\rangle |e \rangle$ state is embedded into the continuum of particle-hole states $a_{d_i}^\dagger a_h |\Omega\rangle |g \rangle$ and can thereby decay into this continuum with the rate $\Gamma = 2\pi \rho | \langle e| \langle \Omega | V_\mathrm{HFI} | a_{d_i}^\dagger a_{h_r} |\Omega\rangle |g \rangle |^2$. Here the hole $h_r$ is such that the energy difference $\varepsilon_{d_i} - \varepsilon_{h_r}$ matches the nuclear excitation energy $\omega_\mathrm{nuc}$ and $\rho$ is the valence band density of states. The rate has to be summed over all allowed 
final states, including phonon degrees of freedom which bring in Frank-Condon factors in the rate evaluation. This internal conversion process quenches the nuclear transition.
The vibronic couplings further ``float'' the hot resonant hole state $h_r$ to the top of the valence band. Finally, a radiative particle-hole recombination emits a fluorescence photon that is red-shifted compared to $\omega_\mathrm{nuc}$. We find that this model semi-quantitatively explains the observed short-timescale fluorescence spectrum of Fig.~\ref{fig:panel_plot}(b-d) and Fig.~\ref{fig:filter_trans}; full numerical simulations will be reported elsewhere.

This mechanism relies on the energy of the $5f$ defect states $\varepsilon_{d_i}$ being lower than $\omega_\mathrm{nuc}$. 
Thus, the long-timescale signal observed from the nuclear isomeric transition is potentially due to \thor{} atoms that are substituted into the crystal in arrangements that have $\varepsilon_{d_i} > \omega_\mathrm{nuc}$ or vanishing $\rho$ near the isomeric energy. 
Finally, even without coupling to defect states,
the nuclear radiative lifetime may still be shortened by effects of the crystal electric field, which mix Th electronic states of opposite parity. 
The hyperfine interaction couples the nuclear degree of freedom to these mixed-parity states, allowing for a competing $E1$ decay channel for the isomeric state. 
Such effects may complicate efforts to  determine the radiative lifetime of the bare \thor{} isomeric state from that observed in crystalline hosts. 

Given these results and those of Ref.~\cite{Tiedau2024}, it is clear that the \thor\ isomeric transition is now measured with laser spectroscopic precision, completing the journey started by Kroger and Reich~\cite{Kroger1976} nearly 50 years ago. 
With a narrow feature identified in two separate crystalline hosts, a number of exciting experiments and applications can be explored. 
These include studying the variation of the transition properties (frequency and linewidth) with chemical environment~\cite{Tkalya1996}, cooperative nuclear spontaneous emission~\cite{Tkalya2011}, and the construction of what may prove to be the ultimate time keeping device, the nuclear clock~\cite{Rellergert2010a,Campbell2012}. 

\section{Acknowledgement}
The authors thank Yafis Barlas, Arlete Cassanho, Yih-Chung Chang, Hans Jenssen, Saed Mirzadeh, and Cheuk Ng for valuable discussions.
This work was supported by NSF awards PHYS-2013011 and PHY-2207546, and ARO award W911NF-11-1-0369.
ERH acknowledges institutional support by the NSF QLCI Award OMA-2016245.
This work used Bridges-2 at Pittsburgh Supercomputing Center through allocation PHY230110 from the Advanced Cyberinfrastructure Coordination Ecosystem: Services \& Support (ACCESS) program, which is supported by National Science Foundation grants \#2138259, \#2138286, \#2138307, \#2137603, and \#2138296.

\bibliographystyle{apsrev4-2}
\bibliography{ref,Th229-apd,ThoriumSearch,Harry_refs}
%\newpage

\include{SI}

\end{document}

%% file: SI.tex
% \documentclass[aps,prl,onecolumn,superscriptaddress]{revtex4-1}
% \pagestyle{plain}
% \usepackage{graphicx}
% \usepackage{bbold}
% \usepackage{amsmath}
% \usepackage{upgreek}
% \usepackage{xcolor}
% \usepackage[utf8]{inputenc}
% \usepackage{graphicx}
% \usepackage[margin=1in]{geometry}
% \usepackage[dvipsnames]{xcolor}
% \usepackage{braket}
% \usepackage[version=3]{mhchem}

% \newcommand{\thor}{$^{229}$Th}
% \newcommand{\regthor}{$^{232}$Th}
% \newcommand{\lisaf}{LiSrAlF$_6$}

% \begin{document}

\section{Supplementary Information}
\section{Obtaining the Fluorescence Spectrum via Gauassian Deconvolution}
The spectrum of the fluorescence was obtained by measuring its transmission through various filter configurations. 
The number of fluorescent photons (normalized to the amount of excitation photons), $Y_i$ that make it through a particular filter configuration is given by 
\begin{equation}
    Y_i = \int \rho(\lambda)~T_i(\lambda)~\eta_i(\lambda)~d\lambda~,
\end{equation} 
where $\rho(\lambda)$ is the spectrum of the fluorescence, $T_i$ are the filter transmission functions (provided by the manufacturer and measured when possible), and $\eta$ captures the remaining quantum efficiencies of the detector and geometric acceptance of the lens system. 

In order to determine $\rho$, a functional form for the spectrum was chosen, $\rho^*(\lambda,\Vec{\xi} )$, which is determined by a set of parameters $\Vec{\xi}$.
For a particular value of $\Vec{\xi}$ we may then compute the following 
\begin{equation}
    Y^*_i(\Vec{\xi}) = \int \rho^*(\lambda,\Vec{\xi})~T_i(\lambda)~\eta_i(\lambda)~d\lambda~,
\end{equation} and from that compute
\begin{equation}
    \chi^2 = \sum_i \frac{\left( Y_i - Y^*_i(\Vec{\xi})\right)^2}{\sigma_i^2}~,
\end{equation} where the sum runs over all filter configurations, and $\sigma_i$ are the uncertainties in the normalized fluorescence counts. Minimizing $\chi^2$ with respect to the parameters $\Vec{\xi}$ will converge to the true spectrum if $\rho$ is of a functional form described by $\rho^*(\lambda,\Vec{\xi} )$.

Given the observed filter data, it was clear that a major contribution to the signal was a sharply peaked component in the region of 150~nm to 200~nm, with a `redder' background component. 
This was reflected in the improvement of $\chi^2$ from $\approx 600$ with one Gaussian function to $\approx 8$ with the addition of another Gaussian. 
Adding more Gaussians did little to improve the value of $\chi^2$. 
The choice of Gaussians to describe the spectrum is only reflective of their simplicity as peaked functions.
It is likely the true spectrum is not exactly described by our parameterization, and the value quoted by our fit for the "sharp" part of the spectrum should only be seen as a rough estimate.

\section{Isomeric Transition Frequency Systematic Uncertainty}
The chief source of systematic uncertainty in the determination of the isomeric transition frequency is the calibration of the VUV laser frequency.
The VUV frequency is given by the difference mixing relation $\omega = 2\omega_{u} - \omega_v$ and therefore depends on the measurement of $\omega_u$ and $\omega_v$, the UV and visible laser frequency, respectively. 
These frequencies are measured using a Bristol 871B wavemeter and augmented with a calibration via a Xe reference line. 
Specifically, the wavelength of the doubled pulsed dye laser (UV PDL) is calibrated by scanning the two-photon  $5p^{6 ~1}S_0~\rightarrow~5p^5\left(^2P^{\circ}_{3/2} \right) 6p~^2\left[1/2\right]_0$ transition in Xenon and comparing to the value on the NIST Database. 
The peak was always within 0.5 pm of the expected value (499.2576~nm for the fundamental). 
Thus, the UV PDL wavelength was accurate to within $\approx$ 1.2~GHz.
The the visible dye laser (VIS PDL) wavelength was measured by the same wavemeter, whose stated calibration uncertainty is 1~pm, or $\approx$ 500~MHz at $785.7$~nm. 
If we assume the worst case that the errors are anti-correlated then $\sigma_{VUV} = 2\sigma_{u} + \sigma_{v} = 3$~GHz.
radiatively of accompanied by the emission of phonons.
\begin{figure*}[h!]
    \centering
    \includegraphics[width=\textwidth]{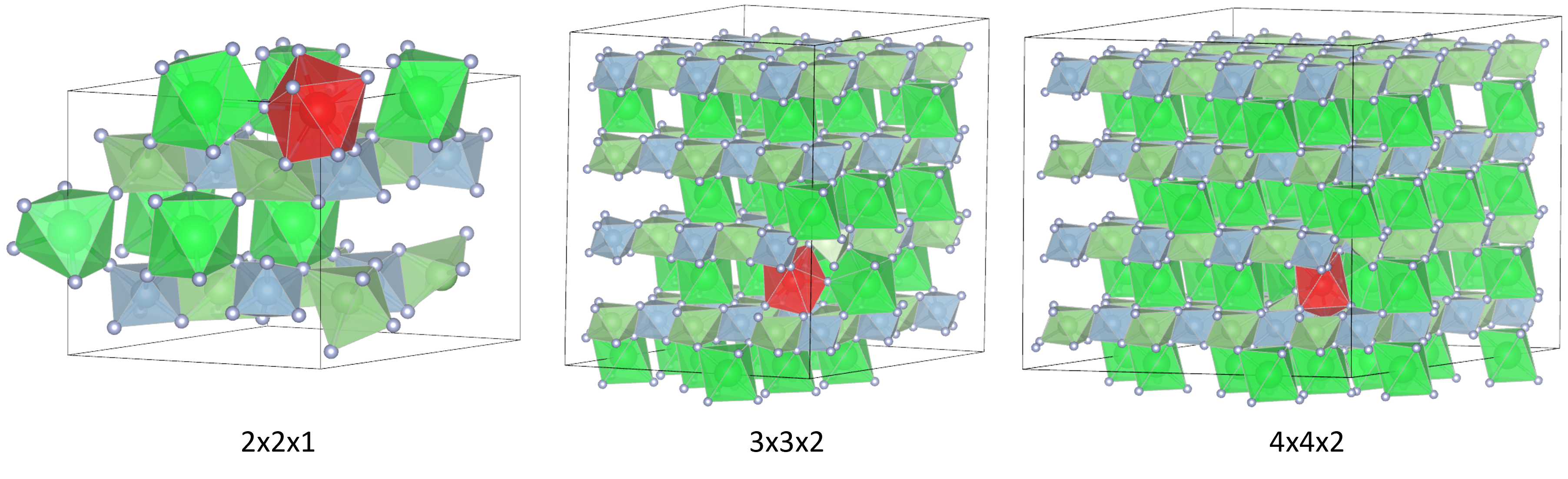}
    \caption{Structures used to model Th:\ce{LiSrAlF6} labelled as $a \times b \times c$ supercells of \ce{LiSrAlF6}. Li is shown in dark green, Sr in light green, Al in blue, F in grey, and Th in red.}
    \label{fig:GM supercells}
\end{figure*}

\section{Density functional theory calculations}
Density Functional Theory (DFT) calculations were used to study the electronic states and energy levels of the thorium atoms in \ce{LiSrAlF6}.
DFT calculations on were performed with VASP\cite{RN12}, version 6.3, using the PAW\cite{RN14} method with a plane-wave cutoff of 500 eV and a spin-restricted formalism.
The PBE\cite{RN13} functional was used for all structural optimizations, and the modified Becke-Johnson (MBJ)\cite{RN489,RN490} functional was used for electronic properties.
Optimizations of $2 \times 2 \times 1$ supercells were done with a 4-4-4 $k$-point grid.

\subsection{Defect structures}
The lowest-energy thorium defects were identified following a procedure developed in a previous study of \ce{Th\text{:}LiCaAlF6}~\cite{RN460}.
Thorium strongly favors the +4 oxidation state, but \ce{LiSrAlF6} contains no +4 cations, exchanging Th for any other atom causes a change in the overall charge.
Charge balancing therefore requires more complex defects, such as \ce{Th^{4+}} replacing \ce{Al^{3+}} with an \ce{Li+} vacancy or \ce{Th^{4+}} replacing \ce{Ca^{2+}} with addition of two interstitial \ce{F-}.
These charges are based on simple chemical considerations and are not imposed in the electronic structure calculations.
The defect structure search in \ce{Th\text{:}LiCaAlF6} began by generating all possible symmetry-inequivalent charge-neutral defects, constrained by a maximum number of atoms which can be added or removed to the structure.
This created 1625 distinct defects in a $2 \times 2 \times 1$ supercell of the host material, 1566 of which were successfully structurally optimized.
For our study of \ce{Th\text{:}LiSrAlF6} we took these 1566 optimized structures of \ce{Th\text{:}LiCaAlF6}, replaced Ca by Sr and set the lattice parameters to those of \ce{LiSrAlF6}, and reoptimized them.
1534 of these optimizations converged successfully.

The resulting structures do not all have the same stoichiometries, and so their energies must be corrected to account for the additional/missing atoms. 
We do this by adding or subtracting the energies of side products required to balance any atoms lost or gained from the \ce{Th\text{:}LiSrAlF6}. These may be the binary fluorides \ce{LiF}, \ce{SrF2}, and \ce{AlF3}, or the ternary products \ce{Li3AlF6} and \ce{SrAlF5} because these are more stable than the relevant combinations of binary fluorides.
With these corrected energies we can compare the stabilities of defects with different stoichiometries.

The lowest-energy defect is thorium replacing a strontium atom with two interstitial fluorine atoms, which is written in Kroger-Vink notation as \ce{Th^{..}_{Sr} + 2F_{i}$'$}.
This contrasts with \ce{Th\text{:}LiCaAlF6} in which the lowest energy defect was found to be Th replacing Al with a neighbouring Li vacancy (\ce{Th^._{Al} + V_{Li}$'$})~\cite{RN460}.
The difference may be because \ce{Sr^{2+}} is larger than \ce{Ca^{2+}} so Sr can more easily accommodate additional fluorides in its coordination sphere.
In this structure Th is coordinated to 8 fluorides in a low-symmetry geometry which may be described as a distorted bicapped octahedron.
The lowest-energy $2 \times 2 \times 1$ structure found by the screening procedure is shown in Figure \ref{fig:GM supercells}.

In fact, the screening of defects produces 30 structures within 0.015 eV of the lowest energy structure, all with the same stoichiometry.
Automatic symmetry detection, using the Atomic Simulation Environment (ase) python package,\cite{RN533} finds that only 7 of these are symmetry-unique.
Visual inspection of these structures and calculation of the average Th-F bond lengths suggest that they are in fact all identical, consistent with their very similar energies.
We therefore use only one of them for further structural and electronic analysis.

A $2 \times 2 \times 1$ supercell may be too small for computation of accurate electronic properties because the effective thorium concentration is $1.05\times10^{21} \text{ cm}^{-3}$ and the minimum distance between periodic images of Th is 10.3 \AA{}.
To reduce unphysical Th-Th interactions, we performed electronic structure calculations on $3 \times 3 \times 2$ and $4 \times 4 \times 2$ supercells, with effective Th concentrations of $2.33\times10^{20} \text{ cm}^{-3}$ and $1.31\times10^{20} \text{ cm}^{-3}$, respectively, and minimum Th-Th distances of 15.5 \AA{} and 20.6 \AA{}, respectively.
These supercells are shown in Figure \ref{fig:GM supercells}.

\subsection{Defect electronic states}
The reason for studying the electronic states of the thorium defects is to determine whether the excited nuclear state of \ce{^{229}Th} can be excited or decay by coupling to an excited electronic state.
DFT typically underestimates band gaps; in pure \ce{LiSrAlF6} the band gap computed by our PBE method is 7.88 eV, significantly below the experimental value of 10.69 eV.
This is a problem because the nuclear excited state lies at approximately 8.35 eV, so PBE predicts that the excited state lies within the conduction band.
This is qualitatively wrong, and prevents understanding of the electronic defect location relative to the conduction band.
Our solution to this is to use the mBJ (modified Becke-Johnson) meta-GGA functional~\cite{RN489,RN490}, which is exceptionally successful at computing band gaps for a fraction of the cost of a hybrid functional.
mBJ gives a band gap for pure \ce{LiSrAlF6} of 11.3 eV, an overestimation of about 5\%, which is important because it is above the nuclear excitation energy so we can place the defect electronic states in the band gap without needing to apply an empirical shift of the conduction band, or ``scissor operator'', as has been done for calculations using the PBE and HSE functionals~\cite{RN531,RN532}.
The other important aspect of mBJ is its low cost-to-accuracy ratio which will allow us to calculate the large supercells described above to minimize spurious Th-Th interactions.
For calculations of electronic properties, such as the density of states and energies of the defect electronic states, we will therefore use the mBJ functional.
Structural optimizations and determination of the relative energies of different defects must still be done with PBE because mBJ is a ``potential only'' functional which gives accurate electronic properties for a given structure but cannot be used to compare different structures~\cite{RN487}.
\begin{figure*}[t!]
    \centering
    \includegraphics[width=\textwidth]{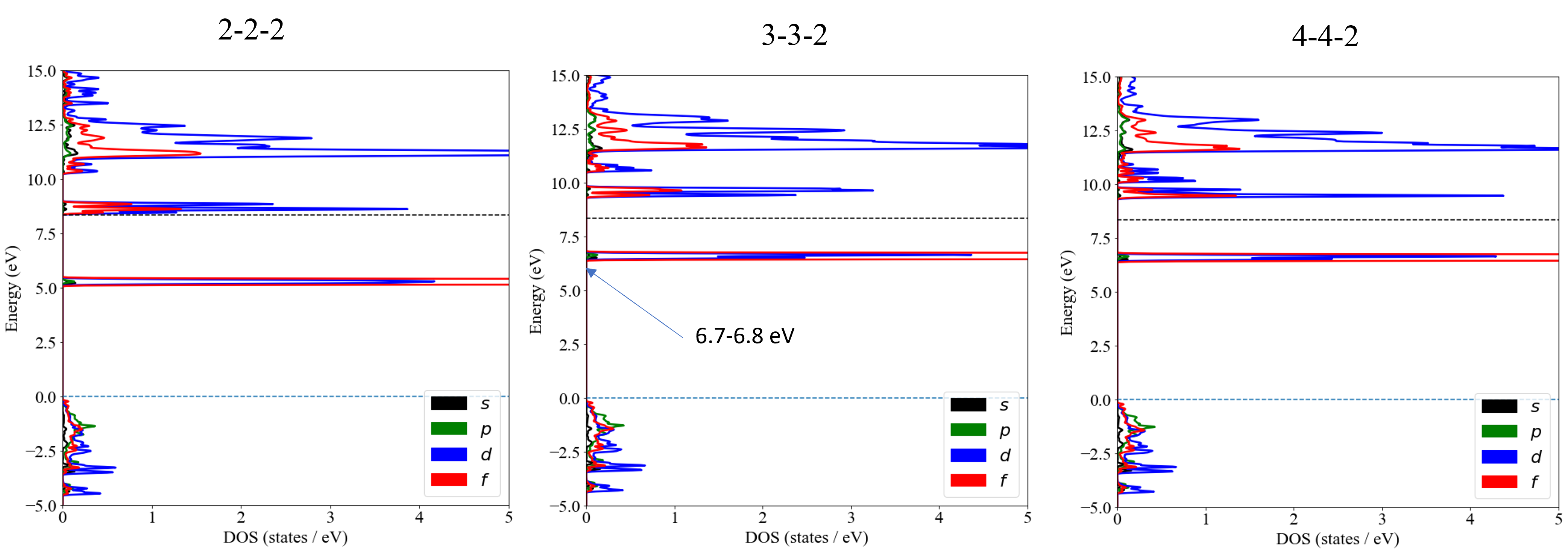}
    \caption{Th PDOS in a 3x3x2 supercell computed with different $k$-point grids. The dashed line denotes the energy of the excited \ce{^{229}Th} nuclear state relative to the top of the valence band.}
    \label{fig:k-point convergence}
\end{figure*}
\begin{figure*}[b!]
    \centering
    \includegraphics[width=0.33\textwidth]{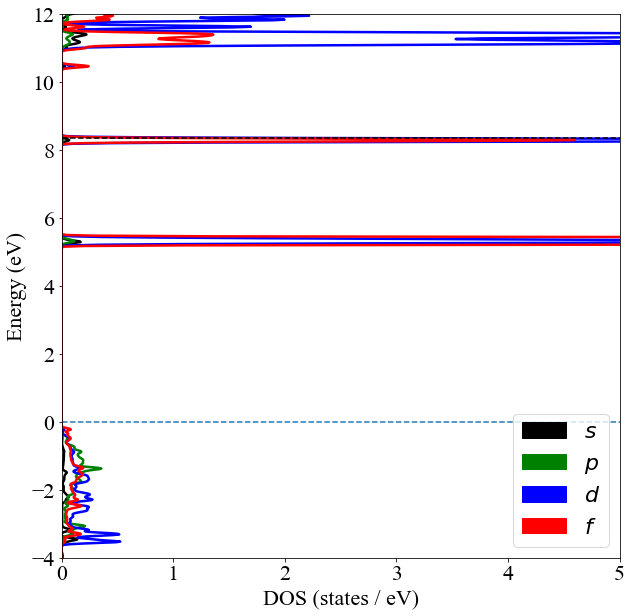}
    \includegraphics[width=0.33\textwidth]{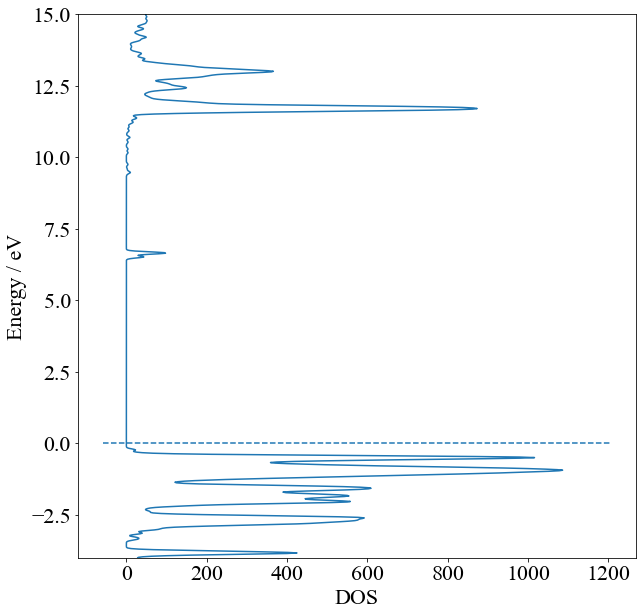}
    \caption{(left) Th PDOS in a 4x4x2 cell computed at the $\Gamma$ point. The dashed line denotes the energy of the excited \ce{^{229}Th} nuclear state relative to the top of the valence band. (right) Total DOS for the 3x3x2 supercell of Th:\ce{LiSrAlF6}.}
    \label{fig:4x4x2 supercell PDOS}
\end{figure*}
% \begin{figure}[b!]
%     \centering
%     \includegraphics[width=0.33\textwidth]{SI/442_kpoints_total_DOS_formatted.png}
%     \caption{}
%     \label{fig:3x3x2 total DOS}
% \end{figure}

We tested the convergence of the thorium projected density of states (PDOS) as a function of the size of the $k$-point grid in the calculation.
The results are shown in Figure \ref{fig:k-point convergence}.
In the 2-2-2 grid the Th $5f$ levels appear just above 5 eV, while in the 3-3-2 and 4-4-2 grids these levels appear at 6.8 eV.
These calculations show that a 3-3-2 $k$-point grid is necessary to converge the DOS in the $3 \times 3 \times 2$ supercell.
Figure \ref{fig:4x4x2 supercell PDOS} shows the thorium PDOS computed in a $4 \times 4 \times 2$ supercell.
The size of this cell meant that we were only able to do a $\Gamma$ point calculation, \textit{i.e.} a $k$-mesh with only a single $k$ point.
The PDOS closely matches the non-converged plot in Figure \ref{fig:k-point convergence} in that the energies of the excited states are underestimated, indicating that the DOS is not converged in this supercell in a $\Gamma$ point calculation.
We therefore use the $3 \times 3 \times 2$ supercell with a 4-4-2 $k$-mesh for all subsequent electronic structure analysis.
The total DOS in Figure \ref{fig:4x4x2 supercell PDOS}(right) shows the defect states arond 7 eV and then the bulk conduction band at around 11.5 eV, in good agreement with the value computed for pure \ce{LiSrAlF6}.

%\bibliography{ref,Th229-apd,ThoriumSearch,SI/Harry_refs}